\documentclass[NETN]{stjour}
\articletype{Research}
\usepackage{rotating}
\usepackage{tabularx}
\usepackage{booktabs}
\usepackage[english]{babel}
\usepackage{varioref}
\usepackage{hyperref}
\usepackage{cleveref}
\usepackage{changepage}
\usepackage{csquotes,booktabs,dcolumn}
\usepackage[title]{appendix}
\usepackage{textpos}

\newcolumntype{d}[1]{D{.}{.}{#1}}
\newcolumntype{Y}{>{\centering\arraybackslash}X}

\begin{document}
\title[A Longitudinal Analysis of University Rankings]{A Longitudinal Analysis of University Rankings}

\author[Author Names]
{Friso Selten\affil{1},
Cameron Neylon\affil{2},
Chun-Kai Huang\affil{2},
Paul Groth\affil{1}}

\affiliation{1}{Informatics Institute, University of Amsterdam, Amsterdam, The Netherlands}
\affiliation{2}{Centre for Culture and Technology, Curtin University, Perth, Australia}

\keywords{University Rankings, Longitudinal Analysis, Comparative Analysis, Principal Component Analysis, Factor Analysis}

\begin{abstract}
Pressured by globalization and the increasing demand for public organisations to be accountable, efficient and transparent, university rankings have become an important tool for assessing the quality of higher education institutions. It is therefore important to carefully assess exactly what these rankings measure. In this paper, the three major global university rankings, The Academic Ranking of World Universities, The Times Higher Education and the Quacquarelli Symonds World University Rankings, are studied. After a description of the ranking methodologies, it is shown that university rankings are stable over time but that there is variation between the three rankings. Furthermore, using Principal Component Analysis and Exploratory Factor Analysis, we show that the variables used to construct the rankings primarily measure two underlying factors: a universities reputation and its research performance. By correlating these factors and plotting regional aggregates of universities on the two factors, differences between the rankings are made visible. Last, we elaborate how the results from these analysis can be viewed in light of often voiced critiques of the ranking process. This indicates that the variables used by the rankings might not capture the concepts they claim to measure. Doing so the study provides evidence of the ambiguous nature of university ranking's quantification of university performance.
\end{abstract}


\section{1 Introduction} \label{sec: introduction}

Over the past 30 years, the public sector has been subject to significant administrative reforms driven by an increased demand for efficiency, effectiveness and accountability. This demand sparked the creation of social measures designed to evaluate the performance of organizations and improve accountability and transparency \citep{romzek2000dynamics}. Universities, as part of this public sector, have also been subject to these reforms \citep{espeland2007rankings}.
One of the measures taken in the higher education domain to serve this need for accountability and transparency is the popularization of university rankings (URs). University rankings are ``lists of certain groupings of institutions [...] comparatively ranked according to a common set of indicators in descending order'' \citep[p.~5]{usher2006world}. 

The idea of comparing universities dates to the 1980’s when the US News \& World Report released the first ranking of American universities and colleges. The process however gained major attention in 2003 with the release of the Shanghai league table \citep{stergiou2014impact}. Many new URs were established since then; the most notable ones: the THE-QS and the Webometrics Ranking of World Universities in 2004, the NTU Ranking and the CWTS Leiden Ranking in 2007. In 2009, the THE and QS rankings split and they publish separate rankings since 2010.
These rankings make it easy to quantify the achievements of universities and compare them to each other. Universities therefore use the rankings to satisfy the public demand for transparency and information \citep{usher2006world}. Furthermore, rankings were met with relief and enthusiasm by policy makers and journalists but also by students and employers. Students use them as a qualification for the value of their diploma, employers to assess the quality of graduates and governments to measure a university's international impact and its contribution to national innovation \citep{hazelkorn2007impact, van2009european}. The internationalisation of higher education has increased the demand for tools to assess the quality of university programs on a global scale \citep{altbach2007internationalization}. Increasing importance of URs can be understood from this perspective because they provide a tool for making cross-country comparisons between institutions.
The impact of rankings is unmistakable. They affect judgments of university leaders and prospective students, but also the decisions made by policy makers and investors \citep{marginson2014university}. For certain politicians, to have their country's universities in the top of the rankings became a goal in itself \citep{saisana2011rickety, billaut2009should}.

University rankings also quickly became subject to criticism \citep{stergiou2014impact}. Fundamental critiques of the rankings are twofold. First, some researchers question whether the indicators used to compute the ranking are actually a good proxy for the quality of a university. It is argued that the indicators the rankings use are not a reflection of the attributes that make up a good university \citep{billaut2009should, huang2012opening}. In addition, researchers reason that URs can become a `self-fulfilling prophecy'; a high rank creates expectations about a university and this causes the university to remain in the top of the rankings. For  example,  prior  rankings  influence  surveys  that  determine future rankings, they influence funding decisions and universities conform their activities to the ranking criteria \citep{espeland2007rankings, marginson2007global}.

The other criticisms focus on the methodologies employed by the rankings. This debate often revolves around the weightings placed on the different indicators that compose a ranking. The amount of weight placed on certain variables is decided by the rankings' designers, but research has shown that making small changes to the weights can cause a major shift in ranking positions. Therefore, the position of a university is largely influenced by decisions made by the rankings' designers \citep{marginson2014university, saisana2011rickety, dehon2009uncovering}. Also, the indicator normalization strategy used when creating the ranking can influence the position of a university \citep{moed2017critical}. Normalisation is thus, next to the assignment of weightings, a direct manner in which the ranking designers influence ranking order. Furthermore, It has been suggested that rankings are biased towards universities in the United States or English speaking universities, e.g. by using a subset of mostly English journals to measure the number of publications and citations \citep{van2005fatal, pusser2013university, vernon2018university}. Last, there is evidence that suggests that there are major deficiencies present in the collection of the ranking data, i.e the data used to construct the rankings is incorrect \citep{van2005fatal}.

The aim of this research is to better understand what it is that university rankings measure. This is studied by examining the data that are used to compose the rankings. We assess longitudinal patterns, observe regional differences and analyze whether there are latent concepts that underly the data used to build the rankings, i.e. if the variables used in the ranking can be categorised into broader concepts. The relation between the result from these analysis to the various criticisms described above will also be discussed. 
Three rankings are analysed in this study; The Academic Ranking of World Universities (ARWU), The Times Higher Education World University Ranking (THE) and the Quacquarelli Symonds World University Rankings (QS). These rankings are selected because they are seen as the most influential and they claim international coverage. They are also all general in that they measure the widest variety of variables as they focus not only on research but also measure teaching quality \citep{scott2013ranking, aguillo2010comparing}.

\subsection{1.1 Related Work}
We take a data driven approach to our analysis, which is somewhat uncommon in the literature. The most notable works that study university rankings using such an approach are \citet{aguillo2010comparing}, \citet{moed2017critical}, \citet{dehon2009uncovering}, \citet{safon2013global}, \citet{soh2015overall} and \citet{docampo2011using}.

\citet{aguillo2010comparing} study the development of the ARWU and the, at that time still publishing a ranking together, THE-QS rankings. This research shows that rankings differ quite extensively from each other but that they do not change much over the years. This is also confirmed by the research of \citet{moed2017critical} which shows that, when analysing five university rankings rankings (besides the ARWU, THE and QS this paper also considers the Leiden and U-Multirank rankings), only 35 universities appear in the top 100 of every rankings. Furthermore, this research examines relations between similar variables that the rankings measure. These analysis prove that citation measures between the different rankings in general are strongly correlated. Also, variables that aim at measuring reputation and teaching quality show moderate to strong correlation \citep{moed2017critical}.
Where \citet{moed2017critical} explores the relation between the ranking variables using correlations, these relations have also been analysed using more sophisticated techniques as Principal Component Analysis and Exploratory Factor Analysis.  \citet{dehon2009uncovering} use this first technique to study the underlying concepts that are measured by university rankings. Their research provides insights in the ARWU ranking by showing that the 2008 edition of this ranking measured two distinct concepts; the volume of publications and the quality of research conducted at the highest level. This is also found by \citet{docampo2011using} who applies Principal Component Analysis on data from the ARWU and shows that the extracted components can be used to assess the performance of a university at a country level. 
\citet{safon2013global} and \citet{soh2015overall} both apply Exploratory Factor Analysis on university rankings. \citet{safon2013global} shows that the ARWU ranking measures a single factor, while the THE ranking measures three distinguishable factors. \citet{soh2015overall} suggests that the ARWU ranking only measures academic performance, while the THE and QS rankings also include non-academic performance indicators. 

We take inspiration from this prior work but move beyond it by performing our analysis longitudinally, over three rankings, using multiple analysis approaches as well as performing geographic and sample comparisons. 

\noindent Specifically, the contribution of this paper is fourfold:
\begin{enumerate}
\item It describes the evolution of, and gives a comparison between, the three major university rankings over the past seven years.

\item It shows the results of a multi-year robust Principal Component and Factor Analysis of the university ranking data, expanding on the work of \citet{dehon2009uncovering, safon2013global, soh2015overall}.

\item It provides evidence that university rankings are primarily measuring two concepts and discusses the implications of this finding.

\item It demonstrates a new visualisation of how the position of specific (groups of) universities in the rankings changes over time.
\end{enumerate}

The structure of this paper is as follows; first, a general explanation of the rankings methodologies and data collection is given in the Sections 2 and 3. Then, our exploratory analysis on the ranking data is discussed Section 4. This section also studies longitudinal stability and cross-ranking similarity. This is followed by the presentation of our analysis of the latent concepts underlying the rankings using Principal Component and Factor Analysis (Section 5). Finally, implications of the results and limitations of the study are discussed.

\section{2 Ranking Methodologies} \label{sec: ranking_methodology}
This section briefly outlines what concepts the rankings use to compute ranking scores and the variables they use to evaluate these concepts. In the next three sections, after each concept the weight assigned to this concept when calculating a universities overall ranking score is indicated between parenthesis.

\subsection{2.1 ARWU Ranking}
The Academy Ranking of World Universities aims to measure four main concepts: Quality of Education (Alumni, 0.1), Quality of Faculty (Award, 0.2; HiCi, 0.2), Research Output (NS, 0.2; PUB, 0.2) and Per Capita Performance (PCP, 0.1). Quality of education is operationalized by counting the number of university graduates that have won a Nobel Prize or Fields Medal. Awards won since 1911 are taken into account, but less value is assigned to prizes that were won longer ago. Quality of Faculty is measured by counting these same prizes, but starting in 1921 and won by university staff. In addition the number of staff members that are listed on the Highly Cited Researchers list composed by Claritive Analytics is used  as an input variable. Research output is measured using the number of papers published in Nature and Science and the total number of papers indexed in the Science Citation Index-Expanded and Social Science Citation Index. The per capita performance variable is a construct of the other five measured variables and – depending on the country a university is in – this construct is either divided by the size of the academic staff to correct for the size of a university, or it is a weighted average of the five other variables. \citep{ARWUmethods}.
For a more in-depth overview of the ARWU methodology see the official ARWU website (\href{http://www.shanghairanking.com}{http://www.shanghairanking.com}), or the articles from \citet{billaut2009should}, \citet{dehon2009uncovering}, \citet{docampo2011using}, \citet{vernon2018university} and \citet{marginson2014university}. 

\subsection{2.2 THE Ranking}
The Times World Ranking of Universities is constructed from the evaluation of five different concepts. Teaching (0.3), Research (0.3), Citations (0.3), International Outlook (0.075) and Industry Income (0.025). Half of the Teaching indicator is constructed using a survey that aims to measure the perceived prestige of institutions in teaching. The other half is made up by the staff-to-student ratio, doctorate-to-bachelor’s ratio, doctorates-awarded-to-academic-staff ratio and institutional income. Research is mostly measured using a survey that seeks to determine a university's reputation for research excellence among its peers. Furthermore, research income and research productivity (the number of publications) are taken into account when constructing the Research variable. Citations are measured by averaging the number of times a university’s published work is cited. This citation measure is therefore normalized for the size of a university. Furthermore, data is normalized to correct for differences in citation rates, e.g. in the life sciences and natural science average citations are much higher than in other research areas such as arts and humanities. Data on citations has been provided by Elsevier using the Scopus database since 2015. Prior to 2015 this information was supplied by Web of Science. International Outlook is measured by evaluating the proportion of international students and international staff and the amount of international collaborations. Industry Income is measured by assessing the research income an institution earns from industry \citep{THEmethods}.
For a more in-depth overview of the THE methodology see the official THE website (\href{https://www.timeshighereducation.com}{https://www.timeshighereducation.com}), or the articles of \citet{vernon2018university} and \citet{marginson2014university}

\subsection{2.3 QS Ranking}
The Quacquarelli Symonds World University Ranking evaluates six different concepts: Academic Reputation (0.4), Employer Reputation (0.1), Faculty/Student Ratio (0.2), Citations per faculty (0.2), International Faculty Ratio (0.05) and International Student Ratio (0.05). Academic Reputation is based on a survey of 80,000 individuals who work in the higher education domain. Employer Reputation is measured by surveying 40,000 employers. The Faculty Student Ratio variable measures the number of students per teacher and is used as a proxy to asses teaching quality. Citations are measured over a five year window using Elsevier’s Scopus database and the score is normalized for the size of a university. As in the THE ranking, since 2015 citations are normalized within each faculty to account for differences in citations rates between research areas. International Faculty and Student ratios subsequently measure the ratio of international staff and ratio of international students \citep{QSmethods}. 
For a more in-depth overview of the QS methodology see the official QS website (\href{https://www.topuniversities.com}{https://www.topuniversities.com}), or articles from \citet{huang2012opening}, \citet{docampo2011using}, \citet{vernon2018university} and \citet{marginson2014university}.

The three rankings all claim to measure three main concepts; quality of education, quality of research and research output. Next to these overlapping concepts the rankings also have unique characteristics. Noticeable is the inclusion of internationality in the THE and QS ranking. This is absent from the ARWU ranking. Also, the THE is the only ranking to include a university's income from industry. 
In general, it can be stated that the methodologies of the THE and QS ranking are quite similar. They use comparable concepts for assessing the quality of a university and similar methodologies for measuring them. The ARWU ranking, while measuring the same concepts, uses different variables and input data to operationalize these concepts. 

\section{3 Data Collection} \label{sec: data collection}
Data for this study has been collected from the official websites of the three university rankings. Data has been retrieved for all variables that compose the rankings described in the previous Section by scraping the university ranking websites.

This research focuses on the ranking years 2012 to 2018 because for these years it was possible to obtain data from the website of all selected rankings. This ensures that for all years analyzed, official data about all three rankings is available. Table \ref{table:N_universities} shows the number of universities present in the rankings per year. The lambda ($\lambda$) column shows the number of universities present in each respective ranking for which all data measured by the rankings is available. The last column (All) shows the number of institutions that are present in all three rankings in that specific year.

\begin{table}[h]
\centering
\caption{Number of universities measured per year}
\label{table:N_universities}
\begin{tabularx}{\columnwidth}{lYYYYYYY}
\toprule
    Year & ARWU & $\lambda$ & THES & $\lambda$ & QS & $\lambda$ & All \\ 
    \midrule
    2012 & 500 & 500 & 400 & 364 & 869 & 392 & 324 \\
    2013 & 500 & 498 & 400 & 367 & 903 & 400 & 326\\
    2014 & 500 & 497 & 401 & 381 & 888 & 395 & 326 \\
    2015 & 500 & 498 & 800 & 763 & 918 & 96 & 413 \\
    2016 & 500 & 497 & 981 & 981 & 936 & 140 & 405 \\
    2017 & 500 & 497 & 1103 & 1103 & 980 & 129 & 414 \\
    2018 & 500 & 497 & 1258 & 1258 & 1021 & 498 & 419 \\
\bottomrule
\end{tabularx}
\end{table}

Different rankings use different names for universities and also within a ranking over the years name changes were observed. Subsequently, to compare universities between years and rankings it was necessary to link all universities to their associated Global Research Identifier Database entry (GRID) \citep{digital-science_2019}. Records were linked using data retrieved from Wikidata \citep{Wikidata}. Wikidata includes the ID's that are assigned by the three rankings for many universities alongside the related GRID. By linking an institution's unique ranking id to the Wikidata database and extracting the relevant GRID, it was possible to match almost all universities. This linkage proved effective; manual inspection of several universities did not detect mismatches. A small number of missing GRIDs were linked by hand. 

\section{4 Exploratory Analysis} \label{sec: exploratory_analysis}
A comparison of the changes in overall positions of universities' rankings is now presented. Two distinct aspects are assessed; changes in the rankings over time and the dissimilarities of the three rankings in the same year with respect to each other.

Three different measurements are used to evaluate these relationships. First, the number of overlapping universities (O), i.e the number of universities that are present in both rankings. Second, Spearman rank correlation coefficient (F), this coefficient measures the strength of the association between overlapping universities \citep{gravetter2016statistics}. To also assess the relationship between rankings including non-overlapping universities a third test, the inverse rank measure (M), as formulated by \citet{bar2007some}, is calculated. This test is also used to compare rankings in the research of \citet{aguillo2010comparing}. The M-measure assesses ranking similarity while factoring in the effect of non-overlapping universities. This is accomplished by assigning non-overlapping elements to the lowest rank position + 1. In the case of two URs with size k, if a university appears in ranking A but does not appear in ranking B, then the university is assigned to rank k+1 in ranking B. The M-measure subsequently calculates a normalized difference between the two rankings \citep{aguillo2010comparing}. Resulting M-scores should be interpreted as follows; below 0.2 can be considered weak similarity, between 0.2 and 0.4 low similarity, values between 0.4 and 0.7 medium similarity, between 0.7 and 0.9 as high similarity and above 0.9 as very high similarity \citep{bar2007some}. Some universities were assigned the same position in the rankings because of a tie in score. These universities were assigned to the mid position, i.e. two universities that are ranked fifth are both assigned to place 5.5.

\subsection{4.1 Longitudinal Ranking Stability}\label{sec: longitutional_paterns}
First, changes within rankings over the past seven years are reviewed. Because the ARWU ranking assigns a singular rank only to universities in the the top 100 of the ranking. A comparison where more institutions are taken into account can be found in Section A of the supplementary material \citep[see][]{friso_selten_2019_3594326}. In general, the results of this analysis do indicate that rankings are also stable beyond the top 100. However as is explained there, these results should be interpreted with care. Within the top 100, the three measurements (Overlap, Spearmans correlation coefficient and the M-measure) can be found in Table \ref{table:Cross-similarity}. This table shows for each ranking the years from 2013 to 2018, as indicated in the left column. For each of these years, each ranking is compared to the data for the years 2012 to 2017 (first row) of the same ranking.

\begin{textblock}{0.89}[0,0](-0.27,0.02)
\begin{table*}[h]
\centering
\caption{Similarity between ranking years (O: Overlap; F: Spearmans correlation coefficient, M: M-measure)}
\label{table:Cross-similarity}
\begin{tabular*}{\textwidth}{l@{\extracolsep{\fill}}@{}lccccccccccccccccc}

\toprule
    
    Measure &
    \multicolumn{3}{c}{2012} &
    \multicolumn{3}{c}{2013} &
    \multicolumn{3}{c}{2014} &
    \multicolumn{3}{c}{2015} &
    \multicolumn{3}{c}{2016} &
    \multicolumn{3}{c}{2017} \\
    \cmidrule(lr){2-4}
    \cmidrule(lr){5-7}
    \cmidrule(lr){8-10}
    \cmidrule(lr){11-13}
    \cmidrule(lr){14-16}
    \cmidrule(lr){17-19}

    &
    O & F & M & O & F & M & O & F & M & O & F & M & O & F & M & O & F & M \\
    \midrule
    
    {\textbf{ARWU}} & & & & & & & & & \\
    2013 & 
    98 & 1.0 & 0.97   \\
    2014 & 
    92 & 0.98 & 0.97 &
    93 & 0.98 & 0.96   \\
    2015 & 
    93 & 0.97 & 0.97 &
    93 & 0.98 & 0.95 &
    97 & 0.99 & 0.99  \\
    2016 & 
    89 & 0.91 & 0.91 &
    89 & 0.93 & 0.93 &
    92 & 0.95 & 0.93 &
    91 & 0.97 & 0.93 &  \\
    2017 & 
    87 & 0.91 & 0.9 &
    87 & 0.91 & 0.9 &
    90 & 0.94 & 0.92 &
    89 & 0.96 & 0.92 &
    95 & 0.97 & 0.94 &  & &  \\
    2018 & 
    85 & 0.89 & 0.89 &
    85 & 0.89 & 0.90 &
    87 & 0.93 & 0.91 &
    86 & 0.95 & 0.91 &
    94 & 0.96 & 0.93 &
    95 & 0.97 & 0.97 \\ [1.5ex]
    
    {\textbf{THE}} & & & & & & & & & \\
    2013 & 
    92 & 0.98 & 0.93   \\
    2014 & 
    91 & 0.96 & 0.89 &
    93 & 0.98 & 0.94   \\
    2015 & 
    83 & 0.88 & 0.89 &
    86 & 0.88 & 0.86 &
    85 & 0.90 & 0.85   \\
    2016 & 
    85 & 0.90 & 0.78 &
    87 & 0.90 & 0.75 &
    85 & 0.92 & 0.74 &
    89 & 0.97 & 0.84   \\
    2017 & 
    85 & 0.91 & 0.71 &
    86 & 0.89 & 0.7 &
    85 & 0.91 & 0.68 &
    89 & 0.96 & 0.77 &
    96 & 0.98 & 0.91  \\
    2018 & 
    85 & 0.91 & 0.69 &
    82 & 0.90 & 0.67 &
    81 & 0.92 & 0.66 &
    86 & 0.95 & 0.76 &
    91 & 0.97 & 0.88 &
    94 & 0.98 & 0.95 \\ [1.5ex]
     
    {\textbf{QS}} & & & & & & & & & \\
    2013 & 
    95 & 0.98 & 0.91  \\
    2014 & 
    94 & 0.96 & 0.88 &
    97 & 0.98 & 0.9   \\
    2015 & 
    93 & 0.90 & 0.81 &
    93 & 0.91 & 0.87 &
    93 & 0.92 & 0.82   \\
    2016 & 
    91 & 0.90 & 0.8 &
    90 & 0.90 & 0.82 &
    90 & 0.91 & 0.81 &
    95 & 0.98 & 0.93  \\
    2017 & 
    87 & 0.88 & 0.79 &
    87 & 0.89 & 0.8 &
    87 & 0.90 & 0.79 &
    91 & 0.97 & 0.9 &
    94 & 0.99 & 0.96 \\
    2018 & 
    86 & 0.88 & 0.77 &
    84 & 0.88 & 0.78 &
    84 & 0.89 & 0.77 &
    89 & 0.96 & 0.87 &
    92 & 0.98 & 0.93 &
    93 & 0.98 & 0.95 \\
\bottomrule
\multicolumn{15}{l}{Note: All Spearmans correlations (F) were significant: $p<.001$} \\
\end{tabular*}
\end{table*}

\end{textblock}
\clearpage

These analyses show that all three rankings are stable over time. A large portion of universities are overlapping for every year. Furthermore, all Spearman correlations coefficients are significant with large effect sizes. This signifies there is not much change in the ranking positions of overlapping universities between years. 
The M-measures provides some more insights in the changes of the rankings over time.
For the ARWU ranking this measurement shows strong similarities. Even when comparing the ranking from 2012 with the one from 2018 the M-measure is very high and only 15 universities do not overlap. 
The THE ranking is more volatile. For example, similarities from the 2018 and 2017 ranking with the ones from 2012, 2013 and 2014 are less strong. This may be connected with the shift from using Web of Science to Scopus as a source of citation data between 2014 and 2015, and if so be indicative of a sensitivity to data sources. 

However, when considering that the number of universities ranked by the THE ranking is three times higher in 2018 than in these earlier years the relationship between them is still quite high. Also, the M-measure between consecutive years shows strong similarities. The change that is present, is thus subtle.
The QS ranking is also very similar over the years. Consecutive years show very high similarity. But the latest ranking also shows high similarity with all previous years, with an M score of 0.77 indicating high similarity when comparing the 2012 ranking with the one from 2018. 

Overall our conclusion is that the rankings are very stable over time. The top 100 institutions of all rankings are significantly correlated between all years and also the M-measure shows very strong similarities between most years. From the three rankings the THE showed most, although subtle, change over time; it is the only ranking in which the M-measure showed a medium similarity between some years.
From these results the conclusion can be drawn that universities in the top 100 are largely fixed. There are not many new institutions that enter and subsequently few institutions that drop out the top 100. Additionally, within the top 100 of each ranking there is little change in position between years. This stability can be explained by the fact that the rankings use rolling averages to measure publications and citations. Furthermore, the ARWU ranking includes prizes won since 1911. In all rankings, subsequent ranking years are thus partly based on the same data. Furthermore, the fact that it is hard to move positions in the top of the rankings, despite the differences between them is consistent with the idea  that the rankings may be having a circular effect, reinforcing the positions that universities hold. This effect is likely to be strongest in the THE and QS rankings because these use reputation surveys, which are likely to be influenced by previous rankings. However, research by \citet{safon2019inter} shows previous rankings might also influence research performance, indicating there might an additional circular effect present in the ARWU ranking.

\subsection{4.2 Similarity Between Rankings}\label{sec: cross-ranking similarity}
Next, we review the similarity between rankings. For each year, the three rankings are compared to each other.
The same three measurements are used to test these relationships. However, as well as analyzing the top 100 universities, the similarity between the top 50 and 50 to 100 range of the rankings are independently examined. The results of this analysis can be found in Table \ref{table:Within-similarity}. Comparisons are shown for each year analysed between the THE and ARWU, and between the QS and ARWU, and the QS and THE rankings. We observe no large discrepancies between years in how similar the rankings are with respect to each other. This is as expected since each ranking does not change much over time. 

There is however a difference between the top 50 and positions 51 to 100. The overlap measurement in the top 50 of each ranking shows that 60 to 70 percent of the universities overlap between rankings. The ranks of these overlapping universities are also significantly correlated. However, the M-measure shows medium similarity, caused by the relatively high amount of non-overlapping universities. In the 50 to a 100 range the similarity between the rankings is very weak. Not even half of the universities overlap and the correlations between the rankings in all years, but the correlation between the ARWU and the THE rankings in 2013, are not significant. The M-measure also shows weak to very weak similarities between rankings in this range. In the top 100 the THE and ARWU rankings and THE and QS rankings overlap with more than seventy universities. Between the ARWU and QS there is a bit less overlap. This also results in an M-measure that is lower than that between the QS and THE and the ARWU and THE rankings. However, all M-measures can be classified as being of medium strength. Furthermore, the Spearman correlation is significant for all comparisons for the top 50 and top 100. The M-measure indicates more similarity at the top hundred than at the top 50 level. This is caused by the fact that the M-measure assigns more importance to the top of the rankings and when comparing the top 100 range there are less non-overlapping universities, i.e. universities that are in the top 50 of one ranking but not in the top 50 of the other ranking, are likely to be in the top hundred of the other ranking.

\begin{textblock}{0.89}[0,0](-0.27,0,02)
\begin{table}[h]
\centering
\caption{Similarity between different rankings (O: Overlap; F: Spearman correlation coefficient; M: M-measure)}
\label{table:Within-similarity}
\begin{tabular}{llccccccccccccccccc}

\toprule
    
    Measure &
    \multicolumn{6}{c}{Top 50} & 
    \multicolumn{6}{c}{50 - 100} & 
    \multicolumn{6}{c}{Top 100} \\
    \cmidrule(lr){2-7}
    \cmidrule(lr){8-13}
    \cmidrule(lr){14-19}

    &
    \multicolumn{3}{c}{ARWU} &
    \multicolumn{3}{c}{THE} &
    \multicolumn{3}{c}{ARWU} &
    \multicolumn{3}{c}{THE} &
    \multicolumn{3}{c}{ARWU} &
    \multicolumn{3}{c}{THE} \\
    \cmidrule(lr){2-4}
    \cmidrule(lr){5-7}
    \cmidrule(lr){8-10}
    \cmidrule(lr){11-13}
    \cmidrule(lr){14-16}
    \cmidrule(lr){17-19}

    &
    O & F & M & O & F & M & O & F & M & O & F & M & O & F & M & O & F & M \\
    \midrule
    
    {\textbf{2012}} & & & & & & & & & \\
    THE & 
     37 & 0.83*** & 0.57
     & & & & 
     19 & 0.22 & 0.13 
     & & & & 
     70 & 0.81*** & 0.59 & & &  \\
    QS & 
     28 & 0.62*** & 0.48 & 
     36 & 0.77*** & 0.49 & 
     16 & 0.0 & 0.17 & 
     15 & -0.27 & 0.05 & 
     64 & 0.68*** & 0.51 & 
     74 & 0.74*** & 0.53  \\ [1.5ex]
    
    {\textbf{2013}} & & & & & & & & & \\
    THE & 
     35 & 0.82*** & 0.58
     & & & & 
     21 & 0.46* & 0.13
     & & & & 
     73 & 0.83*** & 0.60 & & &  \\
    QS & 
     29 & 0.66*** & 0.51 & 
     36 & 0.73*** & 0.52 & 
     15 & 0.28 & 0.10 & 
     14 & -0.02 & 0.08 & 
     63 & 0.69*** & 0.53 & 
     72 & 0.72*** & 0.56 \\ [1.5ex]
     
    {\textbf{2014}} & & & & & & & & & \\
    THE & 
     38 & 0.83*** & 0.60 
     & & & & 
     22 & -0.19 & 0.15 
     & & & & 
     71 & 0.78*** & 0.62 & & &  \\
    QS & 
     32 & 0.72*** & 0.48 & 
     36 & 0.70*** & 0.50 & 
     16 & 0.22 & 0.16 & 
     15 & 0.06 & 0.21 & 
     62 & 0.65*** & 0.50 & 
     72 & 0.71*** & 0.54 \\ [1.5ex]
     
    {\textbf{2015}} & & & & & & & & & \\
    THE & 
     35 & 0.75*** & 0.51 
     & & & & 
     20 & 0.24 & 0.18
     & & & & 
     68 & 0.79*** & 0.54 & & &  \\
    QS & 
     29 & 0.7*** & 0.57 & 
     35 & 0.80*** & 0.55 & 
     16 & 0.34 & 0.09 & 
     17 & 0.34 & 0.13 & 
     61 & 0.70*** & 0.57 & 
     76 & 0.72*** & 0.58 \\ [1.5ex]
     
    {\textbf{2016}} & & & & & & & & & \\
    THE & 
     37 & 0.83*** & 0.54
     & & & & 
     21 & -0.35 & 0.12
     & & & & 
     74 & 0.71*** & 0.57 & & &  \\
    QS & 
     31 & 0.67*** & 0.58 & 
     38 & 0.77*** & 0.56 & 
     17 & -0.21 & 0.12 & 
     17 & 0.2 & 0.13 & 
     66 & 0.63*** & 0.58 & 
     76 & 0.75*** & 0.59  \\ [1.5ex]
     
    {\textbf{2017}} & & & & & & & & & \\
    THE & 
     36 & 0.81*** & 0.55 
     & & & & 
     20 & -0.1 & 0.09
     & & & & 
     73 & 0.75*** & 0.57 & & &  \\
    QS & 
     30 & 0.66*** & 0.57 & 
     37 & 0.73*** & 0.55 & 
     13 & 0.43 & 0.05 & 
     13 & -0.2 & 0.07 & 
     63 & 0.64*** & 0.58 & 
     72 & 0.75*** & 0.58  \\ [1.5ex]
     
    {\textbf{2018}} & & & & & & & & & \\
    THE & 
     36 & 0.83*** & 0.56
     & & & & 
     22 & -0.3 & 0.11
     & & & & 
     75 & 0.73*** & 0.59 & & &  \\
    QS & 
     31 & 0.68*** & 0.56 & 
     40 & 0.78*** & 0.56 & 
     14 & 0.14 & 0.08 & 
     17 & 0.3 & 0.12 & 
     64 & 0.64*** & 0.57 & 
     74 & 0.8*** & 0.6  \\
\bottomrule
\multicolumn{16}{l}{Note: * $p<.05$, ** $p<.01$, *** $p<.001$} \\
\end{tabular}
\end{table}
\end{textblock}
\clearpage

In general, the top 50 and top 100 between all rankings are quite similar. The M-measure points out a medium relationships, but the correlations between the ranks of overlapping universities are strong and significant. The 50 to a 100 range displays much more differences between the rankings. Not even half of the universities are overlapping, the ranks of overlapping universities are not significantly correlated and the M-measure showed very weak similarity. Finally, no two rankings were clearly more similar to each other than to one other ranking. In Section B of the supplementary material \citep[see][]{friso_selten_2019_3594326} we show a similar analysis for the top 400 institutions. These results need to be interpreted with care but show that in the top 400 there is also quite strong similarity between rankings. At the same time however, a rather large number non-overlapping of institutions is present resulting in medium M-scores.
\section{5 Factor Extraction}\label{sec: Exploring_factors}
The three rankings use overlapping concepts (quality of education, quality of research, research performance) but diverse input variables to evaluate these concepts. The above findings show that the rankings do not vary much over time but that the similarity between rankings is less and differs according to the ranking position analysed. We now take a more in-depth look at the input measures.

Previous research suggest that there are two latent factors underlying the ARWU ranking of 2008 and three underlying the THE ranking in 2013 \citep{dehon2009uncovering, safon2013global}. To further examine the similarities and differences between rankings, we analyze if these factors are stable in the rankings over time. This was analyzed using two techniques; Principal Component Analysis (PCA) which has been employed by \citet{dehon2009uncovering} and \citet{docampo2011using} and Exploratory Factor Analysis (EFA) as used by \citet{safon2013global} and \citet{soh2015overall}. The studies of \citet{dehon2009uncovering} and \citet{safon2013global} only reviewed a subset of the ranking data by studying the top 150 or a group of overlapping universities. We are interested in comparing the overall structure of the rankings over multiple years. Therefore, all universities present in the rankings are analyzed, only universities for which the rankings do not provide information on all input measures are removed, because PCA and EFA cannot be applied on missing values. The number of universities that were analyzed each year can thus be seen in the lambda columns in Table \ref{table:N_universities}. 
All input measures analyzed were scaled to have unit variance.

While PCA and EFA are related and often produce the same results, the application of both techniques has two advantages. First, the university ranking data show multivariate outliers. Results from both the PCA and EFA will be influenced by this. Therefore, for both analyses robust techniques are implemented. By applying two methods we can have more confidence that the extracted factors are genuine. Furthermore, PCA and EFA measure different relationships. PCA describes how much of the total variance present in the data can be explained by the extracted components. EFA tries to explain the correlations between variables and only considers shared variance \citep{osborne2008best}. Therefore when observing correlated variables using EFA that together explain a substantial part of the variance as indicated by PCA, there is a strong indications that the input measures are related to a latent concept.

\subsection{5.1 Principal Component Analysis} 
PCA is implemented using a robust method as formulated by \citet{hubert2005robpca} using the rrcov package for R \citep{todorovrobust}. All R and Python scripts used to perform all analyses in this study can be found at \citet{friso_selten_2019_3594326}. This method is robust against the influence that outliers will have on the PCA results. The loadings of the PCA were oblique rotated, because the analyzed variables are expected to be correlated and to make the results easier to interpret. To confirm that this method produces sensible results, the robust PCA method was tested on the top 150 universities from the ARWU ranking of 2008 in an effort to reproduce the results from the analysis of \citet{dehon2009uncovering}. Using this PCA method a comparable loading structure to that of Dehon et al. was found, namely that the ARWU consists of two components where the first component is loaded by the Alumni and Award variables and the second component by the other three variables (NS, HiCi \& Award). This confirms the method we use in this study is comparable to that used by \citet{dehon2009uncovering}.

The same analysis was carried out for all years from 2012 to 2018 and for all three rankings. For each ranking the results of the analysis on the 2018 rankings are described in depth. Following this the changes in structure in the other years with respect to the structure observed for 2018 are discussed (the loading structures for these years can be found in Section C of the supplementary material \citep[see][]{friso_selten_2019_3594326}. Because the components can change order we are not referring to them numerically but name them component A, B and C. Furthermore, for the ARWU ranking we decided to remove the PCP variable from the analysis because of how this variable is constructed: it does not measure a separate concept but is a composite of the other ARWU variables and applies a correction for the size of a university. However, this correction is only performed for universities in specific countries (see \citet{ARWUmethods} for a list of these countries). For universities outside these countries, the PCP variable measures a weighted score of the other five variables. Removing this variable for these reasons is common because interpretation of the variable is not feasible; it does not measure a single concept and has different meanings for universities in different countries \citep{dehon2009uncovering, docampo2011using}. 

A rule often used in PCA is to keep only components with eigenvalues exceeding one, because this indicates the component explains more variation than a single variable \citep{kaiser1960application}. Extracting eigenvalues for the ARWU ranking proved this was only true for the first principal component. However, this rule does not always provide a good estimate for the amount of components to retain \citep{cliff1988eigenvalues, velicer1990component}. Inspection of the scree plots, prior research and assessment of the results when keeping one and two components justified extracting the first two components from this ranking \citep{dehon2009uncovering}. For the THE ranking, the first two components had an eigenvalue higher than one, and for the QS ranking the first three components had an eigenvalue exceeding one. Scree plots and analysis of the results confirmed that extracting two and three principal components respectively was justified.
Results of this analysis for the 2018 ranking data can be seen in Table \ref{table:rotated_pca_loadings}.

\begin{table}[h]
\caption{Rotated PCA Loadings on Components 2018}
\centering
\label{table:rotated_pca_loadings}
\begin{tabularx}{\columnwidth}{lYYY}
\toprule
    Measure & PC-A & PC-B & PC-C \\ \midrule
    \textbf{ARWU} & & & \\
    1. Alumni  &  0.04  &  \textbf{0.66}  &   \\ 
    2. Award   & -0.03  &  \textbf{0.48}  &  \\ 
    3. HiCi    & \textbf{-0.66}  & -0.02  &   \\ 
    4. NS      & -0.36  &  0.35  &   \\ 
    5. PUB     & \textbf{-0.84}  & -0.01  &  \\ [1.5ex]
    
    \textbf{THE} & & & \\
    1. Teaching               & -0.28  & \textbf{-0.45}  &  \\ 
    2. Research               & \textbf{-0.42}  & \textbf{-0.45}  &  \\ 
    3. Citations              & \textbf{-0.89}  & -0.12  &   \\
    4. Industry Income         &  0.07  & \textbf{-0.88}  &   \\ 
    5. International Outlook  & \textbf{-0.96}  &  0.12  & \\ [1.5ex]
    
    \textbf{QS} & & & \\
    1. Academic reputation    & \textbf{-0.99}  & -0.05  &  0.03   \\
    2. Employer reputation    & \textbf{-0.92}  &  0.04  &  0.10  \\
    3. Faculty Student        & -0.15  &  0.04  &  \textbf{0.92}   \\
    4. International Faculty  &  0.01  &  \textbf{0.94}  & -0.08  \\
    5. International Student  &  0.02  &  \textbf{0.95}  &  0.10   \\
    6. Citations              & \textbf{-0.55}  &  0.14  & \textbf{-0.58}  \\
\bottomrule
\multicolumn{4}{l}{Note: Loadings larger than .40 are in bold} \\
\end{tabularx}
\end{table}

These results show a clear structure in the ARWU ranking. The Alumni and Award variables represent component A and the HiCi and PUB variables component B. The NS variable loads on both. 

This structure is also observed in the years 2016 and 2017. In the years 2012, 2013, 2014 and 2015 the Alumni, Award, NS and HiCi variables load on one component, while only the Pub variable loads on the other component.

For the THE ranking we also observe two components. One input variable (Research) loads on both components, while the other four variables load very distinctively on one of the two components. The Research variable loads components A and B. Component A is also influenced by the Citations and International Outlook variables. Component B gets additionally influenced by the Teaching and Industry Income variables. This structure is also observed in the years 2016 and 2017. Before 2016 there is variability in the loading structure. In the years 2012, 2013 and 2014 the Teaching and Research variables load very strongly together on component A and the International Outlook and Industry Income variables load on component B. 
Citations load on both components. The year 2015 is divergent from the other years, in this year the Citation and International Outlook variables influence component B, and Industry Income explains a big proportion of the variance in the other component. Teaching and Research in that year load on both components.

For the QS ranking, a more clear distinction between components can be observed. The Academic and Employer Reputation variables represent component A. International Faculty and Students represent component B. Last, the Faculty Student and Citations variables form component C. The QS ranking also showed the most stability over time. The first components A and B are the same in all years analyzed. However, the Faculty student variable in 2016 also loads on component A. The Citation variable is most volatile and loads differently across years.

For each of the three rankings, the Robust Principal Component Analysis showed that it is possible to reveal structure in the data. Some variables are stable and load on the same component in all years. However, there are also variables that show more variation.  

\subsection{5.2 Exploratory Factor Analysis}
To explore the factorial structure of the data further an exploratory factor analysis using oblique rotations was performed. First, for all three rankings in all years the Kaiser-Meyer-Olkin measure (KMO) has to be verified to test sampling adequacy, and Bartlett’s test of sphericity ($\chi^2$) needs to be performed to analyze if the correlation structure of the data is adequate for factor analyses. 

The tests indicate that all years of all ranking are adequate for factor analysis. For ARWU in all years $KMO > 0.80$ and Barlett's  $\chi^2$ test is significant ($p < 0.001$). For THE in all years $KMO > 0.55$ and Barlett's  $\chi^2$ test is significant ($p < 0.001$).
For all years of the QS $KMO > 0.52$ and $\chi^2$ test is significant ($p < 0.001$).
The KMO values for the THE and QS ranking are quite low. This indicates the existence of relatively high partial correlations between the variables in these two rankings \citep{pett2003making, field2013discovering}. This shows that there is less unique variance in the THE and QS ranking compared to the ARWU. The existence of high partial correlation in URs is to be expected. The ranking variables attempt to measure university performance it is therefore not surprising that the ranking variables, at least partly, account for common variance. Higher KMO values for the ARWU indicate less partial correlations exists in this ranking. The variables in the ARWU, thus, capture more unique variance. Here, it should be noted that KMO assumes normally distributed data and the ranking data diverts from this. It is useful to test the KMO statistic, but one should not place too much emphasis on this test. That being said, given that this research is of an exploratory nature and in all years the KMO values exceed a minimum 0.50 it is possible to perform factor analysis on the data for all years of all three the rankings \citep{kaiser1974index, hair2014multivariate, field2013discovering}. 

The principal axis factors (PAF) extraction method was used because the data deviates from multivariate normality. Principal axis factoring is the preferred extraction method in this situation \citep{osborne2008best}. The non-iterated version was used because the iterated solution yielded Heywood cases, a common problem when using the iterated version of this method \citep{habing2003exploratory}. The same number of factors were extracted as the number of extracted components in the Principal Component Analysis. Scree tests are also a viable strategy for determining the number of factors to retain in factor analysis and a parallel analysis supported the amount of factors to extract. The results of this analysis for the 2018 ranking data can be found in Table \ref{table:nipa_loadings}, for the other years see Section D in the supplementary material \citep[][]{friso_selten_2019_3594326}.

\begin{table}[h]
\centering
\caption{NIPA Loadings on Factors in 2018}
\label{table:nipa_loadings}
\begin{tabularx}{\columnwidth}{lYYY}
\toprule
    Measure & PA-A & PA-B & PA-C \\ \midrule
    \textbf{ARWU} & & & \\
    1. Alumni & \textbf{0.84} & 0.00 & \\
    2. Award & \textbf{0.86} & 0.01 & \\
    3. HiCi &  0.01 & \textbf{0.79} & \\
    4. NS & \textbf{0.41} & \textbf{0.58} & \\
    5. PUB &  -0.09 & \textbf{0.75} & \\ [1.5ex]
    \textbf{THE} & & & \\
    1. Teaching & \textbf{0.92} & -0.02 & \\
    2. Research & \textbf{0.86} & 0.15 & \\
    3. Citations &  0.16 & \textbf{0.66} & \\
    4. Industry Income & \textbf{0.63} & -0.20 & \\ 
    5. International Outlook & -0.04 & \textbf{0.73 } & \\ [1.5ex]
    \textbf{QS} & & & \\
    1. Academic reputation & \textbf{0.88} & -0.06 & 0.07  \\
    2. Employer reputation & \textbf{0.83} & 0.08 & -0.08  \\
    3. Faculty Student & 0.24 & -0.01 & \textbf{-0.40}  \\
    4. International Faculty & -0.02 & \textbf{0.75} & 0.07   \\
    5. International Students & 0.02 & \textbf{0.76} & -0.06  \\
    6. Citations &  0.26 & 0.13 & \textbf{0.44} \\
\bottomrule
\multicolumn{4}{l}{Note: Loadings larger than .40 are in bold} \\
\end{tabularx}
\end{table}

These results generally follow the results obtained with the PCA,  with the structure being more clear. The ARWU consist of two distinct factors, factor A is loaded by the Alumni and Award variables and factor B strongly by the HiCi, NS and PUB variables. This structure is visible in all years.
The THE ranking is also made up of two factors. Factor A is loaded by the Teaching, Research and Industry Income variables. Whereas factor B is constructed of the Citations and International Outlook variables. This structure is visible in 2015, 2016 and 2017. In 2012, 2013 and 2014 factor A does not get loaded by the Industry Income variable. Factor B in those years is only loaded on by the Citations variable and not by International Outlook.
The QS ranking is made up of three factors. Factor A is loaded by the Academic and Employer Reputation variables, factor B by the International Faculty and Students variables, and factor C by the Faculty Student and Citations variables. The QS ranking shows more volatility than the other two rankings. In all years analyzed Factor A and B are respectively loaded on by the reputation variables and the two variables measuring internationality, but there is variation in how the Faculty Student and Citations variables load. In the years 2012, 2013 and 2014, factor C was loaded on only by the Citations variable while the Faculty Student variable did not load on any of the factors. In 2015 and 2016, both the Faculty Student and Citations variables loaded on factor A together with the Academic and Employer Reputation variables. In 2017, both Citations and Faculty student did not load higher then 0.4 on either of the three factors. 

\subsection{5.3 Explaining the Factors}
The structure in the three rankings was evaluated using two different methods: robust Principal Component Analysis and Exploratory Factor Analysis. The first method is robust against the presence of outliers in the data, while the second is resistant against the data being non-normally distributed. We now examine if the factors that were empirically found by these two analysis, are also theoretically explainable and what underlying concepts these factors measure.

In the ARWU ranking two distinct factors can be observed in the EFA while the PCA shows more volatility. Generally, however, it can be stated that the HiCi, PUB and N\&S variables appear to form a factor together and the Alumni and Award variables form a second factor. This structure was also found in the research of \citet{soh2015overall} and \citet{dehon2009uncovering}. The first factor measures the number of citations and publications and together is weighted 60 percent on the ranking. The variables that form the second factor, Alumni and Award, measure the number of Nobel and Field prizes won by a universities employees or alumni and is weighted 30 percent in the ARWU ranking. 
\citet{safon2013global} came to a different conclusion, showing that all ARWU variables load on the same factor. This study however used a specific subset of the data which had a significant effect on the extracted structure.

In the THE ranking, two distinct factors also are extracted in both the PCA and EFA. The first factor is composed of the Teaching and Research variables. These two variables are measured by multiple sub-variables, as described in the \nameref{sec: ranking_methodology} Section. Only in the years 2016 to 2018 we see this reflected in the results of the robust PCA. In these years the research variable loads on both components. This maybe a reflection that this variable, when correcting for the influence of outliers, is derived from two quite different notional indicators. However when assessing all years and the EFA results, we in accordance with the interpretation of \citet{moed2017critical}, expect the teaching and research variables loading together is caused by the influence of the surveys while the other variables used to construct these variables have little impact because of the low weights assigned to them. This component is therefore mainly a representation of a university's reputation and accounts for 60 percent of the ranking. The second component is, when considering all years, influenced mainly by the Citations variable which provides 30 percent of the final ranking. There is quite some variation in how the Industry Income and International Outlook variables load. These are not clearly related to a single factor, also both only weigh five percent on the ranking. The research of \citet{safon2013global} and \citet{soh2015overall} shows comparable results. However, in these studies the Citations variable loaded with the Research and Teaching variable. Our results suggest that, when taking the whole ranking into account over multiple years, the Citation measure is a separate factor in the THE ranking.

The QS ranking is the only ranking for which the extraction of three factors proved useful according to scree plots and parallel analysis. However, when considering multiple years only two are consistent. The Academic and Employer Reputation variables load together in both PCA and EFA. This suggests, as in the THE ranking that they are a measure of the general reputation of a university. This factor provides 50 percent of the ranking. Also, the International Faulty and International Students variables form a construct together. This factor accounts for 15 percent of the weight in the ranking. The last extracted factor in the QS ranking was not consistent. Both Citations and Student to Staff ratio thus appear to be separate components in this ranking when analyzing multiple years of the QS ranking. They both provide 20 percent. These results differ quite a bit from those obtained by \citet{soh2015overall}, which might be caused by the fact that that study only extracted two factors.

Reviewing these results and assessing what the variables that form the concepts measure shows that in all three rankings in all years there are two overlapping underlying concepts that contribute substantially to the rankings: 1. reputation and 2. research performance. 

In the ARWU ranking, we observed that the N\&S, HiCi and PUB variables often load together. These variables are all proxies for the research performance of a university. The second component is composed of the Alumni and Award variables. Both these variables measure the same achievements but in different groups and can be seen as a proxy for (or influencer of, as indicated by the work of \citep{Philip2012globalization}, a university's reputation.
In the THE ranking, reputation is measured by the Teaching and Research variables, while the Citations variable is measuring research performance. In the QS ranking, Academic and Employer Reputation compose the reputation factor whereas research performance is measured by the Citations variable. 

Also, some non-overlapping concepts were found. PCA and EFA showed that internationality is a separate concept in both the THE and QS ranking while in the ARWU this concept is not represented. Also, in the QS ranking the student-to-staff ratio plays quite an important role. In the other two rankings, this concept is not assigned much importance.

When taking the weights assigned to the variables into account 90 percent of the ARWU ranking, 85 of the THE ranking and 70 percent of the QS ranking is accounted for by the two concepts. Reputation and research performance, are thus very influential in all three rankings. A last difference that can be observed in the rankings is that in the ARWU ranking indicators of research performance are more important, while in the THE and QS rankings the indicators associated with reputation are the most influential. 

\subsection{5.4 Reliability of the Concepts}
The analysis described above concluded that there are two overlapping concepts, 1. reputation and 2. research performance, that represent most of the weight in each of the rankings. In the ARWU ranking, both concepts are a combination of multiple variables. In the THE and QS, only the reputation performance measurement is a multi-item concept. To confirm that the variables that measure one concept together form a reliable scale, the internal validity of the scales was verified. The Spearman-Brown split-half reliability test was used for this because some concepts are composed of two variables \citep{eisinga2013reliability}. The results of these test can be found in Table \ref{table:Spearman_Brown}. They confirm that in all years the scales are internally reliable. This supports the assertion that for all three rankings the factors that consist of multiple variables are reliable scales measuring the same concept across the multiple years. Furthermore, for the THE and QS ranking it can be observed that these scales are more internally reliable when compared to internal reliability for the whole ranking tested using Cronbach's alpha (see supplementary material Section E \citep[see][]{friso_selten_2019_3594326}). While in the ARWU reliability of the scales is comparable to the internal consistency of the complete ranking. This indicates that, while our analysis shows the existence of two internal reliable scales in the ARWU ranking, these concepts are more inter-related than is the case for the THE and QS ranking. This is consistent with the finding that the ARWU ranking is mostly a one-dimensional scale assessing academic performance, while the other two rankings are more multi-dimensional \citep{safon2013global, soh2015overall}.

\begin{table}[h]
\centering
\caption{Spearman Brown scale reliability}
\label{table:Spearman_Brown}
\begin{tabularx}{\columnwidth}{l Y Y Y Y}

\toprule
    
    Scale & 
    \multicolumn{2}{c}{ARWU} & 
    \multicolumn{1}{c}{THE} & 
    \multicolumn{1}{c}{QS} \\
    \cmidrule(l{6pt}r{6pt}){2-3}
    \cmidrule(l{6pt}r{6pt}){4-4}
    \cmidrule(l{6pt}r{6pt}){5-5}

    &
    1 & 2 & 1 & 1 \\
    \midrule
    2012 & 0.86 & 0.88 & 0.95 & 0.82 \\
    2013 & 0.87 & 0.88 & 0.95 & 0.83 \\
    2014 & 0.87 & 0.89 & 0.96 & 0.83 \\
    2015 & 0.87 & 0.88 & 0.95 & 0.88 \\
    2016 & 0.87 & 0.84 & 0.95 & 0.84 \\
    2017 & 0.87 & 0.83 & 0.95 & 0.84 \\
    2018 & 0.87 & 0.83 & 0.95 & 0.89 \\
\bottomrule
\end{tabularx}  
\end{table}

\section{6 Investigating the Scales}
Based on our analysis to this point, we conclude that two concepts underlie all three rankings. To further investigate what these concepts measure, the variables of which they consist were combined. For each ranking, this creates a two dimensional representation of each ranking describing the reputation and research performance of the universities. 

\subsection{6.1 Testing Scale Relationships}
To assess the relationship between these concepts a Spearman correlation test for each year was performed, results can be found in Section F of the supplementary material \citep[see][]{friso_selten_2019_3594326}. These show that all concepts in all years are significantly correlated with each other. Across years the THE and QS reputation performance measurement seem to be correlated most strongly, but also the THE reputation and ARWU research concepts show strong correlation. These differences are however only minor, in general the reputation performance concept of each ranking are not evidently correlated stronger with the reputation performance concept of the other rankings than with the research performance concept of the other rankings and vice versa. 
The relation between concepts is explainable from the circular effect as described in the study of \citep{safon2019inter}. This shows there is a reputation-survey-reputation effect in the THE and QS rankings as well as in the ARWU ranking, even though it does not include reputation surveys. Furthermore, \citet{robinson2019mining} reverse this argument. They hypotheses that the answers people give on surveys is influenced by publication and citation data. 

These two interpretations can explain the results of our study. We identify the existence of two latent concepts in all rankings: reputation and research performance. However, these two latent concepts might be influencing each other. They are both influenced by previous rankings and inter-dependent. This explains why the different components are correlated. 

\begin{figure*}[p]
\centering
\caption{Longitudinal developments per geographical region}
\includegraphics[width=1.00\textwidth]{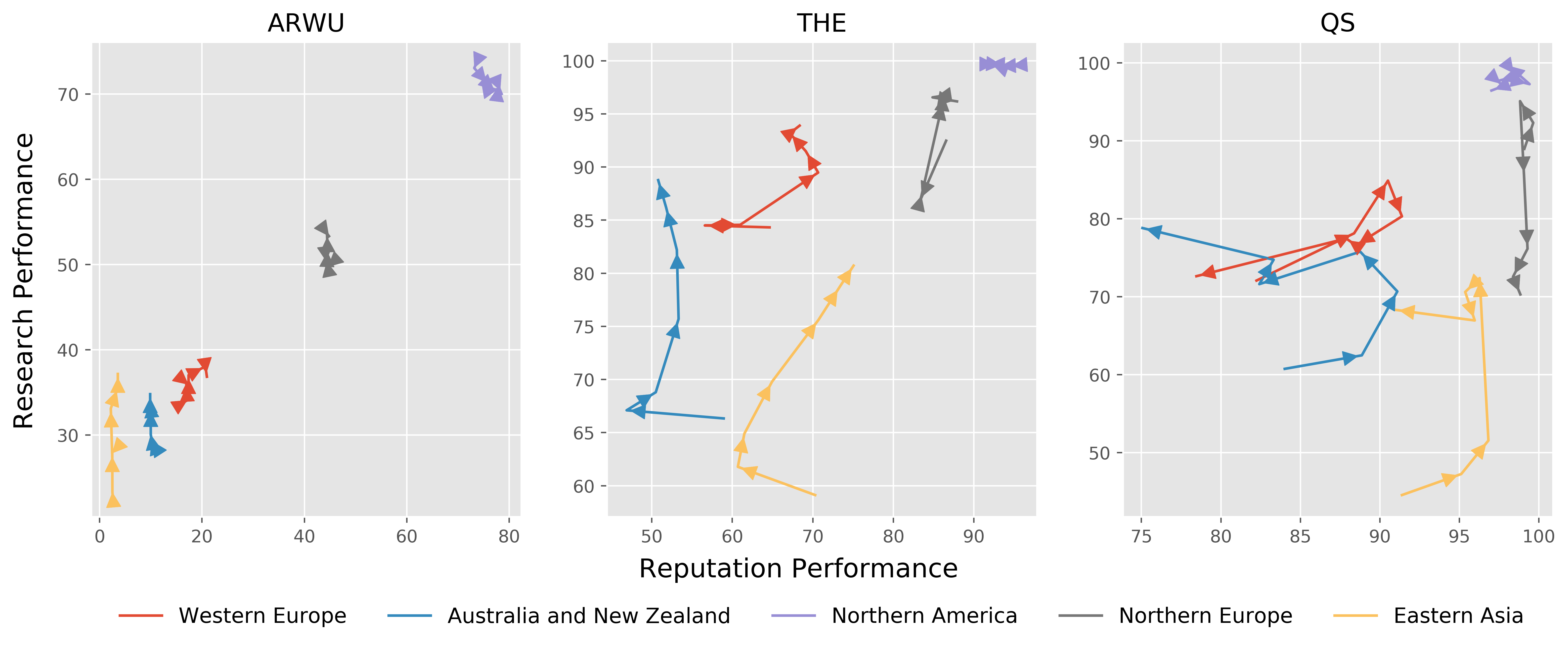}
\label{fig: subregion_ranking}
\caption{Longitudinal developments per language region}
\includegraphics[width=1.00\textwidth]{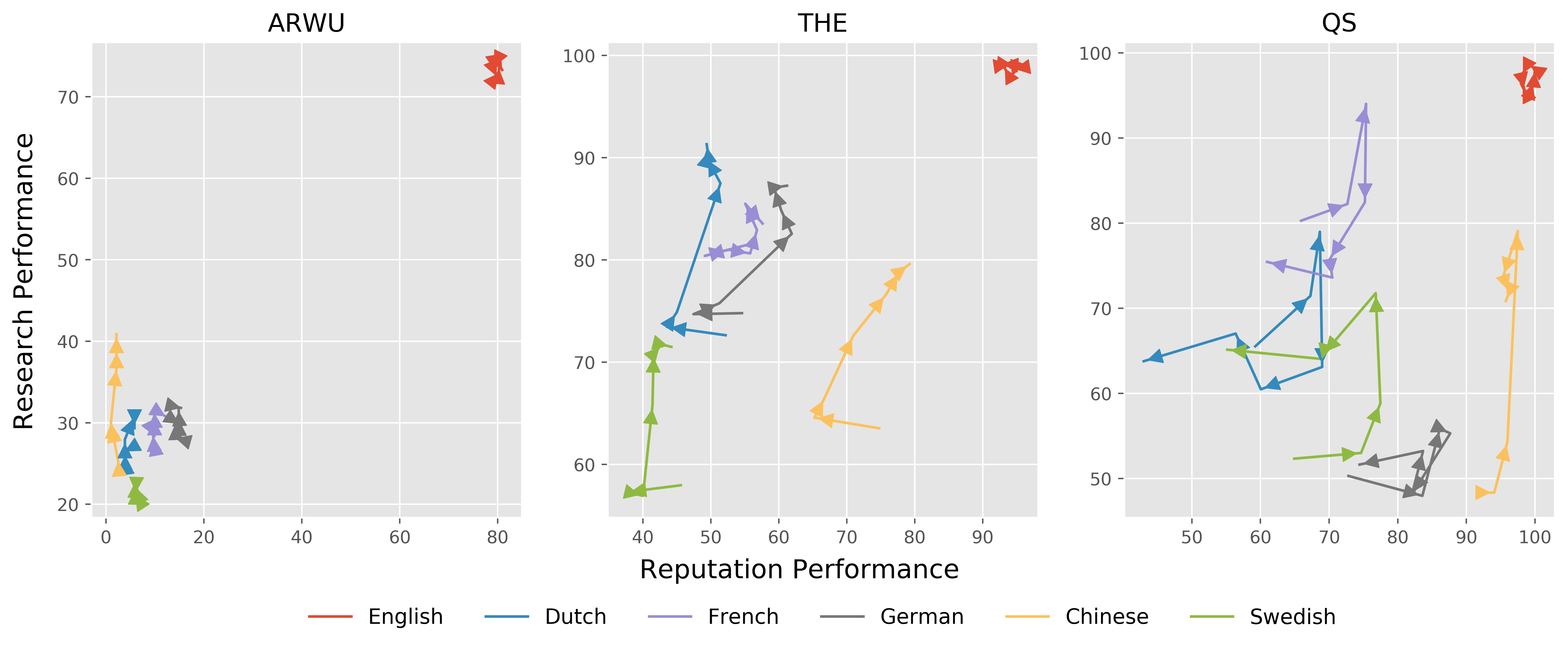}
\label{fig: language}
\caption{Longitudinal developments for a sample of universities}
\includegraphics[width=1.00\textwidth]{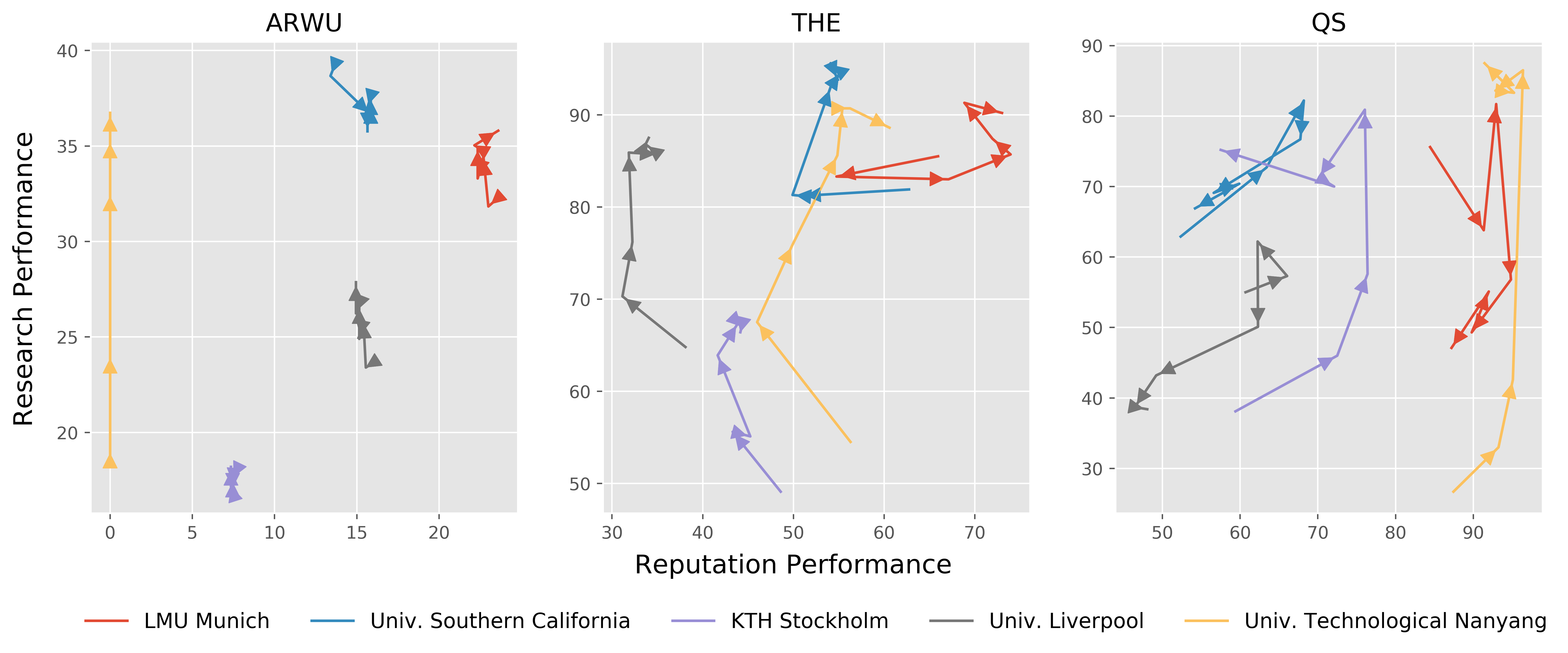}
\label{fig: universities}
The three plots above show the movement of (aggregates) of universities on the two scales; reputation and research performance. Each arrow indicate the data point for a given year, starting in 2012 and ending in 2018, The arrow direction shows the movement from year to year.
\end{figure*}

\subsection{6.2 Plotting the Scales} \label{sec: plots}
The correlation coefficients themselves do not provide much insight into how the different components and factors ('scales' for 
the rest of this discussion) relate to each other. However, they are a two-dimensional reflection of the most important concepts in all three rankings. We were interested in whether using these components as coordinates to map the relationship of outputs and reputations over time for each ranking would provide insight. In particular the question of whether there are differences in the progress made by universities in different regions and by language spoken, whether this could provide evidence for or against claims of bias in the rankings, and if it could provide evidence for or against the circular reinforcement effects discussed above.  

To ease interpretation of the plots, they were created  using a subset of universities that are present in all rankings. This results in a subset of 87 high ranked  universities. Also, there is a difference in number of universities that are ranked per region. This would skew the comparison between regions. Because in the rankings the top institutions are most important, we chose to only aggregate the results of the top five institutions in each region. Arrows in the figures show the direction of the line, starting in 2012 and ending in 2018.

Figure \ref{fig: subregion_ranking} shows how the three rankings behave on a regional level. There are differences in the relative rankings of universities from different regions. We therefore plot the average of the top five institutions from each region. North America, in the ARWU, is very far ahead of all other regions on both the reputation and research performance scale. South-Eastern Asia, Eastern Asia, Western Europe, Australia and New Zealand are all far behind. Northern Europe appears right in the middle. 
The THE and QS ranking both also show that the top institutions in North America perform best on both scales. However, the advantage with respect to the other regions is much smaller. Northern Europe performs second best on both scales in both rankings, but in the THE and especially QS ranking, Asian universities also perform very well on the reputation measurement.

Another interesting observation from this figure is that in all rankings Asian universities are climbing fast on the research performance scale. Finally, universities in Western Europe and Australia and New Zealand in the THE and QS rankings seem to have quite a low reputation score when compared to their score on the research performance scale. While in the ARWU ranking this is the case for institutions in Eastern Asia; performance on the research scale for universities in this area rose quickly, but they continue to lag behind on the reputation measurement. The ARWU shows strikingly lower movement on the reputation scale than do the other rankings, indicating the slow accumulation of prizes compared to the volatility or responsiveness of a survey-based measure.

In the second set of plots (Figure \ref{fig: language}), an aggregate of the top five universities within certain language regions is displayed. This shows in all three rankings that universities in English speaking countries are ahead on both the reputation and research performance scale. Of all three rankings the ARWU shows biggest difference between English speaking countries and the other language regions. 
In the THE ranking on the research performance scale universities in Dutch, French and German speaking countries perform equal and are around twenty points behind English speaking countries. However, on the reputation scale they are substantially further behind. For universities in China, the opposite is the case. They score well on the reputation scale, but are behind on the research performance scale.
The QS ranking shows that institutions in German speaking countries performing quite well on the reputation scale. While Dutch, French and Swedish speaking countries lag behind on this measurement. Chinese institutions increased their performance substantially on the research performance scale over the years. There is, however, no effect of this increase visible on the reputation scale on which they already performed well.

In Figure \ref{fig: universities}, five universities from diverse countries that are all on average ranked in the 50 to 150 range are compared. When comparing the different plots against each other it can be observed that LMU Munich, the University of Southern California and KTH Stockholm perform similar in all rankings. An interesting case is a comparison of the Universities of Nanyang and Liverpool. The first performs very well on the reputation performance scale when this is measured using surveys as in the THE and QS rankings. In the ARWU ranking Nanyang performs poorly on this scale. This difference might be caused by the fact that this institution was established in 1981, hence it has less alumni or university staff that won a Nobel or Field prize. The University of Liverpool in contrast scores very well on the reputation scale in the ARWU. However, seven out of the nine Nobel prizes acquired by the University of Liverpool were won before Nanyang university was founded. This shows how the use of Nobel and Field prizes by the ARWU ranking to measure reputation can favor older institutions.
Also the behaviour of Nanyang university on the research performance scale is noteworthy. In all rankings in 2012 this institution is ranked one of the lowest on this scale when compared to the other four universities in this plot, and in seven years it climbs to be among the top performers. In the QS ranking where the reputation score of this university is also very good, Nanyang university climbs from being ranked position 47 to 12 in this period. This shows that, while the results in Section 4 indicate that the rankings are stable over the years, there are specific universities that manage to rapidly climb to the top of the rankings.

The plots show that the ARWU ranking assigns high scores on both the research performance and reputation scales to institutions in English speaking countries and particularly in the United States and United Kingdom. Asian universities in the ARWU ranking perform the worst on both scales. This in contrast with the other two rankings in which English institutions are also ranked highest but Asian universities are often are among the best performers on the reputation scale. Last, the figures show that on the research performance scale the rankings have more in common than on the reputation scale - there is more variation visible between the plots when comparing the aggregates or universities on the reputation scale.

We see little correlation overall between reputation and output components for any of the groups in any of the rankings. Substantial changes in the output component (both positive and negative) are generally not correlated with similar movements in reputation, even with some delay. The exception to this may be East Asian and Chinese-speaking universities for which there is some correlation between the increasing output component and reputation, primarily in the THE rankings. However, this may also be due to an unexpected confounder. Increasing publications and visibility, and in particular the global discussion of the importance of the increase in volume and quality of Chinese outputs might lead to more researchers from those universities being selected to take part in the survey. This is impossible to assess without detailed longitudinal demographic data on the survey participants.

In general, however, these plots show little evidence for strong relationships between reputation and research performance. This could be consistent with circular reinforcement effects on reputation, where proxy indicators for reputation are largely decoupled from output performance. Overall, examining single universities or groups does not provide evidence for or against circular reinforcement effects. As shown earlier in this paper, there is little change in the rankings. Circular effects are therefore hard to observe, because for most universities performance on the rankings is quite stable.  

\section{7 Discussion} \label{sec: discussion}
Accelerated by the increased demand for accountability and transparency, university rankings have started to play a major role in the assessment of a university's quality. There has been substantial research criticizing these rankings, however only a few studies have performed a longitudinal data driven comparison of university rankings. This research set out to take an in-depth look at the data of the ARWU, THE and QS rankings. Based on this analysis, we draw out five key findings:

\noindent1. \textit{Rankings primarily measure reputation and research performance} \\
\citet{dehon2009uncovering}, \citet{safon2013global} and \citet{soh2015overall} showed that by using Principal Component and Factor Analysis on university ranking data it is possible to reveal structures that underlie these rankings. In this research, these techniques are applied on multiple years of the ARWU, THE and QS ranking. Results of these analyses provide empirical evidence that all three major university rankings are predominantly composed of two concepts; reputation and research performance. Research performance is measured by the rankings using the number of citations and in the ARWU also the number of publications. Reputation is measured in the ARWU by counting the Nobel Prizes and Field Medals won by affiliated university employees and graduates. The THE and QS rankings mainly measure reputation using surveys. The high weights placed on these two concepts by the rankings is problematic. Surveys are a factor that a university has little to no control over and the measurements used to assess research performance are often claimed to be biased \citep{vernon2018university}.

\citet{moed2017critical} shows that individual citation and reputation indicators are strongly correlated. Building upon this, we examined the correlation between reputation and research performance concepts across the rankings. This showed that all concepts are significantly correlated, but correlations within the ``same'' concept across rankings are not stronger than with the divergent concept. There are multiple explanations possible for this absence of a strong correlation between notionally overlapping concepts. First of all, this can potentially be caused by the different types of normalization used by the rankings \citep{moed2017critical}. Second, previous studies have argued that reputation and research performance might influence each other \citep{safon2019inter, robinson2019mining}. A university having a high number of citations can positively affect its reputation and publications written by scholars working at a prestigious university might get cited more often. This is a plausible assertion and the correlations we identify between non-overlapping concepts are consistent with this argument. However, when we directly visualised the relationships between research performance and reputation scales for a range of universities and groups we did not see evidence for this, as can be seen in Section \nameref{sec: plots}. It is nonetheless worthwhile to further explore this effect in future research to gain more insights into the relation between a university's reputation and research performance.

Therefore, we want to explore a different explanation, that the underlying concepts in different rankings are not measuring the same thing. If they were, a clearly stronger correlation between overlapping concepts should be visible. And if it is not reputation or research performance the rankings measure, what information do they actually provide? This uncertainty is problematic considering the influence university rankings have on society \citep{marginson2014university, saisana2011rickety, billaut2009should}. It also leads us into the next point: the complications that arise when measuring reputation.

\noindent2. \textit{Reputation is difficult to measure} \\
Measuring reputation in itself is not unimportant since graduating from or working at a prestigious university can improve a student's or researcher's job prospects \citep{taylor2007international}, even though the relevance of using surveys to rank universities is debated \citep{vernon2018university}. The rankings should therefore look critically at the methodology used to measure this concept.
The THE and QS rankings both use two different surveys to measure reputation. The results from the PCA and EFA showed that in both rankings these surveys are highly related. This suggests that these surveys do not in practice provide information on the (separate) quality of education and research but actually measure a university's general reputation. This then raises the question what people base their judgment on regarding the reputation of a university. It is not unlikely that the rankings themselves play an important role in this, reinforcing the idea that rankings become a self-fulfilling-prophecy \citep{espeland2007rankings, marginson2007global}. The use of Nobel and Field prizes as a substitute might appear more objective. However we have shown that this leads to a bias towards older universities since it includes alumni who graduated since 1911 and prize winners since 1921. This bias is seen in the example of Nanyang Technological University. Furthermore, because the specifically research oriented prizes (Physics, Chemistry, Medicine, Economics and the Fields medal) are focused on the natural and medical sciences, they are biased against the social sciences and humanities. While the Nobel Peace and Literature prizes have been awarded to academics, this is rare. The last humanities academics to win the Nobel Peace Prize were Jody Williams in 1997 and Elie Wiezel in 1986 and the most recent winner of the Literature prize with a substantive academic post is Jean-Marie Gustave Le Clézio (2008). This measure, along with many of the others, therefore, favors science-oriented universities.

Another concern with the reputation measurement in the ARWU and THE rankings is that these rankings use the variables measuring this reputation concept as proxies for a university's education and research quality \citep{ARWUmethods,THEmethods}. The variables loading together and forming a reliable scale reflect the notion that it is doubtful if this these variables are a good representation of these unique qualities. Especially in the THE ranking case, it seems that the reputation surveys have such a big influence on the variables that these are mainly a reputation measurement. For the QS ranking, while the problem is the same, there is at least the merit that it is explicitly noted in the methodology that the rankings is measuring reputation directly \citep{QSmethods}.

\noindent3. \textit{Universities in the US and UK dominate the rankings} \\
The \nameref{sec: plots} Section shows that universities in English speaking countries are ahead of universities in other regions. This seems to support the critique that rankings are regionally biased toward western, especially English speaking, universities \citep{van2005fatal,pusser2013university}. For all rankings, we see a substantial advantage for English speaking universities on the research performance scale, even though more and more universities in non-English speaking countries publish predominantly in English \citep{curry2004multilingual, altbach2013imperial}. However, despite the fact that both the THE and QS ranking employ methodologies to account for the fact that non-English articles receive less citations universities in English speaking countries still lead the rankings. This is supporting evidence that the biases described in \citet{pusser2013university} and \citet{vernon2018university} are still present. 

There is also a strong regional effect between the rankings on the reputation component. Eastern Asian, especially Chinese universities, score highly on the reputation measurement in the THE and QS ranking. Non-English speaking European universities and institutions from Australia and New Zealand perform substantially worse on this scale, even when the output component is the same or higher. Reputation measures for Australian and New Zealand universities appear particularly volatile in the QS ranking. This may indicate that the THE and QS rankings reputation measurements favor Asian universities. This could be due to increasing profile and marketing, more effective gaming of the survey by top East Asian and Chinese universities or some other difference in the methodology. More research is thus needed to draw definitive conclusions on this matter.

\noindent4. \textit{Rankings are stable over time but differ from each other} \\
Our analysis shows that for all three rankings consecutive years of the same ranking are strongly correlated and are very similar according to the M-measure. This is in accordance with results of \citet{aguillo2010comparing}. This means that it is hard to change position within a ranking. This year-to-year similarity can be explained by different choices made by the ranking designers. All three rankings use rolling averages to calculate research performance indicators. Also, 30 percent of the ARWU ranking is constructed by variables that measure prizes won since 1921 and which are therefore very stable. For the THE and QS stability can be explained by the high weighting assigned to reputation surveys. A university's reputation is not likely to substantially change within one, or even a small number, of years. Generally speaking all URs employ conservative methodologies, which results in the rankings being very stable. 
Circular effects between rankings years, as described by \citet{safon2019inter} and \citet{robinson2019mining}, could also results in the rankings being stable. The plots created in this research did not indicate the existence of such effects but the correlation between reputation and research performance can be taken as evidence for the claim made by \citet{safon2019inter} that research performance is also influenced by prior rankings. 

The rankings were also compared to each other. These analysis showed that in the top 50 and top 100 the different rankings are correlated strongly, however the M-measure indicated only medium similarity, showing substantial variation between the rankings. These results are in accordance with the findings of \citet{aguillo2010comparing} and the overlap measurements of \citet{moed2017critical}. Given the stability, one would expect that the rankings would be more similar, thus, it is surprising that they are not. It is even more noteworthy that there is dramatically more similarity in the top 50 than in positions 51-100. This is most likely caused by the fact that performance differences are only minor between lower ranked universities. Designer choices are, as will be shown next, influential for ranking order and become more influential when the differences in performance are becoming smaller.

\noindent5. \textit{Ranking designers influence ranking order} \\
The relative absence of similarity between the three rankings is noteworthy. Several reasons can be used to explain the differences between the rankings. First, the rankings assign different weights to the variables that compose these concepts which, as been shown in multiple studies, has a large effect on a university's ranking position \citep{marginson2014university, saisana2011rickety, dehon2009uncovering}. It should also be noted that there are also non-overlapping measurements which can explain differences between rankings. For example, the QS assigning a substantial weight to the student-staff ratio and decision of the ARWU to not include internationality are the most important. Second, rankings use different methods to normalise their data. The THE and QS correct their research performance measurement for university size, while in the ARWU raw numbers are used. Choices made by the ranking designers for a specific weighting and normalisation schemes are thus important determinants of the final ranking order \citep{moed2017critical}. Perhaps most importantly, our  paper, in agreement with previous work, shows that the majority of the ranking variables are attempts to quantify two specific concepts of university performance. The differences between the rankings is therefore not what they are trying to measure but how they seek to measure it.

Three limitations of this research should be addressed.
First, we concluded that the reputation variables in the THE and QS ranking loading together is caused by the fact that these are both measuring a general reputation concept. However, it is possible that these variables do actually measure distinct reputation properties, but that teaching quality and research quality are extraordinarily highly correlated. While there is a likely connection between teaching and research quality, we are skeptical that a) this correlation would be so high and b) that survey respondents are in a position to distinguish between details of education and research provision, especially in a context where they are being asked about both. Attempts to distinguish between teaching quality and research quality, such as in the UK's Teaching and Research Excellence Framework show low correlation between highly evaluated institutions. It is thus reasonable to expect that their judgment is, at least partially, caused by more general reputation attributes, for example the number of Nobel and Field prizes won \citep{Philip2012globalization}. More research is needed to identify what influences survey respondent's judgment of a university's reputation and how the selection of respondents and questions might influence that. This could be studied by reviewing the questions used to measure the reputation variables and analysis of the raw data collected from these questionnaires. It may also interesting to see how external data sources relate to these measurements. For example, by measuring the impact of a university appearing in popular or social media \citep{priem2010altmetrics}. Our results might be seen as supportive of the INORMS statement that surveys should not form the basis of rankings \citep{inorms2018recommendations}. In any case, greater transparency on the sample selection and questions posed (as well as how they may have changed) would be of value in probing this issue.

Second, in the QS ranking in some years a number of universities had to be removed from the analysis because of missing data elements. However, since the loadings in the PCA and EFA for the QS were similar across the years, we are quite confident that a genuine structure was extracted from this ranking. Nonetheless, the large number of missing values in the QS makes it unclear how the overall ranking score for a wide range of universities was constructed and makes it hard to study and verify the QS data. We would urge the QS ranking to provide more transparency in this area.

Third, there is research that suggest all ARWU variables measure one concept \citep{safon2013global}. Results of our Principal Component Analysis also showed most variables loading on one component in some years and the second component's eigenvalue did not exceed one. Of the three rankings, the ARWU therefore appears to be measuring the most singular concept. This is most likely caused by the fact that there are a substantial number of universities that score very low on the Alumni and Award variables, which in turn is a logical result of how these variables are measured (see Section \nameref{sec: ranking_methodology}). For the institutions that score low on these two variables the ARWU thus only measures academic performance. But, when reviewing the EFA results and previous work by \citet{dehon2009uncovering}, we think it is reasonable that these Alumi and Award variables are actually measuring a distinct factor.

This paper provided a longitudinal comparison between the three major university rankings. It showed that rankings are stable over time but differ significantly from each other. Furthermore, it revealed that the rankings all primarily measure two concepts: reputation and research performance, but it is likely that these concepts are influencing each other. Last, it discussed these findings in light of the critiques that have been raised on university rankings. This provides insights in what university rankings do and do not measure. Our results also show that there is uncertainty surrounding what the rankings' variables exactly quantify. One thing is however certain; it is impossible to measure all aspects of the complex concept that is university performance \citep{van2009european}. Despite this, universities are focusing and restricting their activities on ranking criteria \citep{marginson2014university}. But because it is unclear what the rankings quantify it also unclear what exactly the universities are conforming to. Universities aim to perform well on ambiguous and inconsistent ranking criteria, which at the same time can hinder their performance on activities that are not measured by the rankings. We conclude that universities should be extremely cautious in the use of rankings and rankings data for internal assessment of performance and should not rely on rankings as a measure to drive strategy. A ranking is simply a representation of the ranking data. It does not cover all aspects of a university’s performance but it may also be a poor measure of the aspects it is intended to cover.

\newpage

\bibliography{NETNbibsamp}



\end{document}